\documentclass[aps, showpacs, showkeys,nofootinbib,floatfix]{revtex4}

\usepackage{amssymb}
\usepackage{amsmath}
\usepackage{graphicx}

\begin{document}

\title{Constraints on modified Chaplygin gas from recent observations and a comparison of its status with other models}

\author{Jianbo Lu}
\email{lvjianbo819@163.com}
\author{Lixin Xu}
\author{Jiechao Li}
\author{Baorong Chang}
\author{Yuanxing Gui}
\author{Hongya Liu}

\affiliation{School of Physics and Optoelectronic Technology, Dalian
University of Technology, Dalian, 116024, P. R. China}

\begin{abstract} In this Letter, a modified Chaplygin gas (MCG) model
of unifying dark energy and dark matter with the exotic equation of
state $p_{MCG}=B\rho_{MCG} -\frac A{\rho_{MCG}^\alpha }$ is
constrained from recently observed data: the 182
 Gold SNe Ia, the 3-year WMAP
 and the  SDSS baryon acoustic peak. It is shown that
 the best fit value of  the three
 parameters ($B$,$B_{s}$,$\alpha$) in MCG model are
 (-0.085,0.822,1.724). Furthermore, we find the
best fit $w(z)$ crosses -1 in
 the past and the present best fit value $w(0)=-1.114<-1$, and the
$1\sigma$ confidence level of $w(0)$ is $-0.946\leq
w(0)\leq-1.282$.
 Finally, we find that the MCG model has the smallest
 $\chi^{2}_{min}$ value in all eight given models. According to the Alaike
Information Criterion (AIC) of model selection, we conclude that
 recent observational data support the MCG model as well as other popular models.
\end{abstract}


\maketitle

\section{Introduction}

{\small {~~~}}~The type Ia supernova (SNe Ia) explorations
\cite{[1]}, the cosmic microwave background(CMB) results from WMAP
\cite{[2]} observations, and surveys of galaxies \cite{[3]} all
suggest that the universe is speeding up rather than slowing down.
The accelerated expansion of the present universe is usually
attributed to the fact that dark energy is an exotic component with
negative pressure. Many kinds of dark energy models have already
been constructed such as $\Lambda$CDM \cite{[4]}, quintessence
\cite{[5]}, phantom \cite{[6]}, generalized Chaplygin gas (GCG)
\cite{[7]}, quintom \cite{[8]}, holographic dark energy \cite{[9]},
and so forth.

On the other hand, to remove the dependence of special properties of
extra energy components, a parameterized equation of state (EOS) is
assumed for dark energy. This is also commonly called the
model-independent method. The parameterized EOS of dark energy which
is popularly used in parameter best fit estimations, describes the
possible evolution of dark energy. For example, $w=w_{0}$=const
\cite{[10]}, $w(z)=w_{0}+w_{1}z$ \cite{[11]},
$w(z)=w_{0}+\frac{w_{1}z}{1+z}$ \cite{[12]},
$w(z)=w_{0}+\frac{w_{1}z}{(1+z)^{2}}$ \cite{[13]},  $
w(z)=\frac{1+z}{3}\frac{A_{1}+2A_{2}(1+z)}{X}-1$ (here $X\equiv
A_{1}(1+z)+A_{2}(1+z)^{2}+(1-\Omega_{0m}-A_{1}-A_{2})$) \cite{[14]}.
The parameters $w_{0}$, $w_{1}$, or $A_{1}$, $A_{2}$ are obtained by
the best fit estimations from cosmic observational datasets.

It is well known that the GCG model has been widely used to
interpret the accelerating universe. In the GCG approach, dark
energy and dark matter can be unified by using an exotic equation
of state. Also, a Modified Chaplygin gas (MCG) as a extension of
the generalized Chaplygin gas  model
 has already
been applied to describe the current accelerating expansion of the
universe \cite{[15]} \cite{[16]} \cite{[17]} \cite{[18]}. The
constraint on parameter B in MCG model,
 i.e., the added parameter relative to GCG model, is discussed briefly by using the location of
 the peak of the CMB radiation spectrum in Ref. \cite{[19]}. In this Letter, we study the constraints on
the best fit parameters ($B,B_{s}$,$\alpha$) and EOS in the MCG
model from recently observed data: the latest observations of the
182
 Gold type Ia Supernovae (SNe) \cite{[20]}, the 3-year WMAP CMB shift parameter \cite{[21]}
 and the baryon acoustic oscillation (BAO) peak from Sloan Digital Sky
Surver (SDSS) \cite{[22]}. The result of this study indicates that
 the best fit value of
parameters ($B,B_{s}$,$\alpha$) in MCG model are
(-0.085,0.822,1.724). Furthermore, we find the
best fit $w(z)$ crosses -1 in
 the past and the present best fit value $w(0)=-1.114<-1$, and the
$1\sigma$ confidence level of $w(0)$ is $-0.946\leq w(0)\leq-1.282$.
At last, because the emphasis of the ongoing and forthcoming
research is shifting from estimating specific parameters of the
cosmological model to model selection  \cite{[23]}, it is
interesting to estimate which model for an accelerating universe is
distinguish by statistical analysis of observational datasets out of
a large number of cosmological models. Therefore, by applying the
recent observational data to the Alaike Information Criterion (AIC)
of model selection, we compare the MCG model with other seven
general cosmological models to see which model is better. It is
found that the MCG model has almost the same support from the data
as other popular models. In the Letter, we perform an estimation of
model parameters using a standard minimization procedure based on
the maximum likelihood method.

 The Letter is organized as follows. In section 2, the MCG model is introduced  briefly. In section 3,
the best fit value of parameters ($B,B_{s}$,$\alpha$) in the MCG
model are given from the recent observations of SNe Ia, CMB and BAO,
and we present the evolution of the best fit of $w(z)$ with
$1\sigma$ confidence level with respect to redshift $z$.
 The preferred cosmological
 model is discussed in section 4 according to  the AIC.
  Section 5 is the conclusion.\\

\section{Modified Chaplygin gas model}

~~~~For the modified Chaplygin gas model, the energy density $\rho$
and pressure $p$ are related by the equation of state  \cite{[15]}

\begin{equation}
p_{MCG}=B\rho_{MCG} -\frac A{\rho_{MCG} ^\alpha },\label{1}
\end{equation}
where $A, B,$ and $\alpha$ are parameters in the model.

Considering the FRW cosmology, by using the energy conservation
equation: $d(\rho a^{3})=-pd(a^{3})$, the energy density of MCG can
be derived as  \cite{[18]}
\begin{equation}
\rho _{MCG}=\rho _{0MCG}[B_s+(1-B_s)(1+z)^{3(1+B)(1+\alpha
)}]^{\frac 1{1+\alpha }},\label{2}
\end{equation}
for A $\neq $ -1, where $a$ is the scale factor, B$_s=\frac
A{(1+B)\rho _0^{1+\alpha }}$. In order to unify dark matter and
dark energy for the MCG model, the MCG fluid is decomposed into
two components: the dark energy component and the dark matter
component, i.e., $ \rho _{MCG}=\rho _{de}+\rho _{dm}$,
$p_{MCG}=p_{de}$. Then according to the relation between the
density of dark matter and redshift:
\begin{equation}
\rho _{dm}=\rho _{0dm}(1+z)^3,\label{3}
\end{equation}
the energy density of the dark energy in the MCG model can be
given by
\begin{equation}
\rho _{de}=\rho_{MCG} -\rho _{dm}=\rho
_{0MCG}[B_s+(1-B_s)(1+z)^{3(1+B)(1+\alpha )}]^{\frac 1{1+\alpha
}}-\rho _{0dm}(1+z)^3.\label{4}
\end{equation}

Next, we assume the universe is filled with two components, one is
the MCG component, and the other is baryon matter component, ie., $
\rho _t=\rho _{MCG}+\rho _b.$ The equation of state of dark energy
can be derived as  \cite{[18]}

\begin{equation}
w _{de}=\frac{(1-\Omega _{0b})[B_s+(1-B_s)(1+z)^{3(1+B)(1+\alpha
)}]^{-\frac \alpha {1+\alpha }}[-B_s+B(1-B_s)(1+z)^{3(1+B)(1+\alpha )}]}{%
(1-\Omega _{0b})[B_s+(1-B_s)(1+z)^{3(1+B)(1+\alpha )}]^{\frac
1{1+\alpha }}-\Omega _{0dm}(1+z)^3},\label{5}
\end{equation}
where $\Omega_{0dm}$ and $\Omega_{0b}$ are present values of the
dimensionless dark matter density and baryon matter component.

 Furthermore, in a flat
universe, making use of the Friedmann equation, the Hubble
parameter $H$ can be written as

\begin{equation}
H^2=\frac{8\pi G \rho_{t}}{3}=H_0^2E^{2},\label{6}
\end{equation}
where $E^{2} =$ (1-$\Omega
_{0b}$)[$B_s+(1-B_s)(1+z)^{3(1+B)(1+\alpha )}$]$^{\frac 1{1+\alpha
}}+\Omega _{0b}(1+z)^3$.  $H_{0}$ denotes the present value of the
Hubble parameter. When $B = 0$,
 equation
(\ref{6}) is reduced to the GCG scenario.

In the following section, on the basis of equation (\ref{6}), we
will apply the recently observed data to find the best fit
parameters ($\Omega_{0b},B,B_{s}$,$\alpha$) in MCG model. For
simplicity, we will displace parameters
($\Omega_{0b},B,B_{s}$,$\alpha$)
 with $\theta$ in the following section.\\

\section{The best fit parameters from present cosmological observations }
~~~~Since type Ia Supernovae behave as Excellent Standard Candles,
they can be used to directly measure the expansion rate of the
universe up to high redshifts ($z\geq 1$) for comparison with the
present rate. Therefore, they provide direct information on the
universe$^{,}$s acceleration and constrain the dark energy model.
Theoretical dark energy model parameters are determined by
minimizing the quantity
\begin{equation}
\chi^{2}_{SNe}(H_{0},\theta)=\sum_{i=1}^{N}\frac{(\mu_{obs}(z_{i})
-\mu_{th}(z_{i}))^2}{\sigma^2_{obs;i}},\label{7}
\end{equation}
where $N=182$ for the Gold SNe Ia data  \cite{[20]},
$\sigma^2_{obs;i}$ are errors due to flux uncertainties, intrinsic
dispersion of SNe Ia absolute magnitude and peculiar velocity
dispersion respectively. The theoretical distance modulus $\mu_{th}$
is defined as
\begin{equation}
\mu_{th}(z_{i})\equiv
m_{th}(z_{i})-M=5log_{10}(D_{L}(z))+5log_{10}(\frac{H_{0}^{-1}}{Mpc})+25,\label{8}
\end{equation}
where
\begin{equation}
D_{L}(z)=H_{0}d_{L}(z)=(1+z)\int_{0}^{z}\frac{H_{0}dz^{'}}{H(z^{'};H_{0},\theta)},\label{9}
\end{equation}
$\mu_{obs}$ is given by supernovae dataset, and $d_{L}$ is the
luminosity distance.

The structure of the anisotropies of the cosmic microwave background
radiation depends on two eras in cosmology, i.e., last scattering
and today. They can also be applied to limit the model parameters of
dark energy by using the shift parameter  \cite{[24]},
\begin{equation}
R=\sqrt{\Omega_{0m}}\int_{0}^{z_{rec}}\frac{H_{0}dz^{'}}{H(z^{'};H_{0},\theta)},\label{10}
\end{equation}
where $z_{rec}=1089$ is the redshift of recombination, $\Omega_{0m}$
is present value of the dimensionless matter density, including dark
matter and the baryon matter component. By using the three-year WMAP
data  \cite{[25]}, $R$ can be obtained as  \cite{[26]}
\begin{equation}
R=1.71\pm0.03.\label{11}
\end{equation}
From the CMB constraint, the best fit value of parameters in the
dark energy models can be determined by minimizing
\begin{equation}
\chi^{2}_{CMB}(H_{0},\theta)=\frac{(R(H_{0},\theta)-1.71)^{2}}{0.03^{2}}.\label{12}
\end{equation}

Because the universe has a fraction of baryons, the acoustic
oscillations in the relativistic plasma would be imprinted onto the
late-time power spectrum of the nonrelativistic matter  \cite{[27]}.
Therefore, the acoustic signatures in the large-scale clustering of
galaxies can also serve as a test to constrain models of dark energy
with detection of a peak in the correlation function of luminous red
galaxies in the SDSS  \cite{[22]}. By using the equation
\begin{equation}
A=\sqrt{\Omega_{0m}}E(z_{BAO})^{-1/3}[\frac{1}{z_{BAO}}\int_{0}^{z}\frac{H_{0}dz^{'}}{H(z^{'};H_{0},\theta)}]^{2/3},\label{13}
\end{equation}
and $A=0.469\pm0.017$ measured from the SDSS data, $z_{BAO}=0.35$,
 we can minimize the $\chi^{2}_{BAO}$defined as  \cite{[28]}
\begin{equation}
\chi^{2}_{BAO}(H_{0},\theta)=\frac{(A(z^{'};H_{0},\theta)-0.469)^{2}}{0.017^{2}}.\label{14}
\end{equation}

As one can find that the gravitational clustering in MCG model
presented in Ref. \cite{[17]}, ensures that the observational
datasets from CMB and BAO can be applied to constrain the MCG model.
Hence, we combine these three datasets to minimize the total
likelihood $\chi^{2}_{total}$
\begin{equation}
\chi^{2}_{total}(H_{0},\theta)=\chi^{2}_{SNe}+\chi^{2}_{CMB}+\chi^{2}_{BAO}.\label{15}
\end{equation}

  On the one hand, since we are interested in the model parameters $\theta$, the
$H_{0}$ contained in $\chi^{2}_{total}(H_{0},\theta)$ is a
nuisance parameter and will be marginalized by integrating
 the likelihood
$L(\theta) =\int d H_{0}P(H_{0})\exp$
 $(-\chi^{2}(H_{0},\theta)/2)$,
where $P(H_{0})$ is the prior distribution function of the present
Hubble constant, and a Gaussian prior $H_{0} = 72 \pm
8kmS^{-1}Mpc^{-1}$  \cite{[29]} is adopted in the Letter. We know
that the prior knowledge of cosmological parameter $\Omega_{0b}$ has
been obtained by several other observations, such
 as $\Omega_{0b}h^{2}=0.0214\pm 0.0020$ from the observation of the deuterium to hydrogen ratio
 towards QSO absorption systems  \cite{[30]},
$\Omega_{0b}h^{2}= 0.021\pm 0.003$ from the BOOMERANG data
\cite{[31]} and $\Omega_{0b}h^{2}=  0.022^{+0.004}_{-0.003}$
 from the DASI results \cite{[32]} for the observation of CMB.
 Thus, in order to get the interesting result for the value of $\Omega_{0b}$, we
 treat $\Omega_{0b}$ as a free
parameter with Gaussian prior distribution centered in
$\Omega_{0b}^{ture}$ with spread $\sigma_{\Omega_{0b-prior}}$. And
following Ref. \cite{[33]},
 the "weak" prior for parameter $\Omega_{0b}$  will be used in our analysis,
  i.e., let it have a relative larger variable range $\Omega_{0b}h^{2}=0.0214\pm 0.0060$ \cite{[33]}.
Thus, the $\chi^{2}_{total}(H_{0},\theta)$ in equation (\ref{15})
will be reconstructed as \cite{[34]}
\begin{equation}
\chi^{2}_{total-prior}(\theta)=\chi^{2}_{total}(\theta)
+\frac{(\Omega_{0b}-\Omega_{0b}^{true})^{2}}{\sigma^{2}_{\Omega_{0b}-prior}},\label{16}
\end{equation}
where $\chi^{2}_{total}(\theta)$ denotes the total
$\chi^{2}(\theta)$ obtained without imposing prior knowledge of
$\Omega_{0b}$.

\begin{figure}
  \includegraphics[width=220pt]{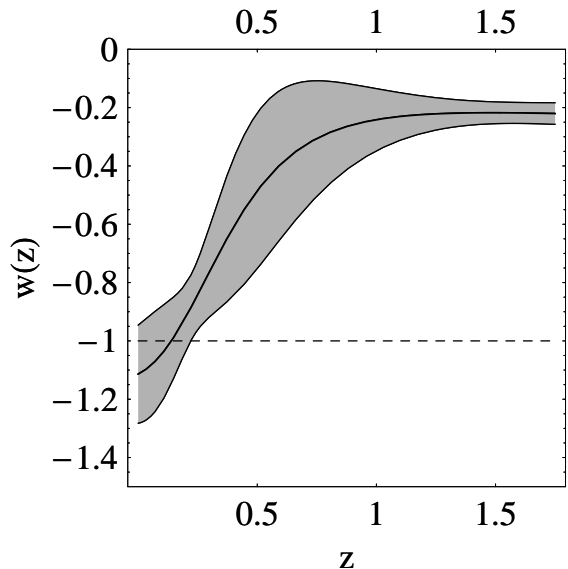}\\
  \caption{The best fits of $w(z)$ with 1$\sigma$ confidence
level (shaded region).}\label{1}
\end{figure}

By using the maximum likelihood method for equation (\ref{16}), we
obtain the best fit values $(\Omega_{0b},B,B_{s},\alpha)$ in the MCG
model (0.041,-0.085,0.822,1.724) with $\chi^{2}_{min}=157.272$.
 Figure 1 shows the $1\sigma$ confidence level of the best fit $w(z)$ calculated by using
 the covariance matrix.
 From Fig.1, it is easy to see that
the best fit $w(z)$ cross -1 at about $z=0.140$ and the present best
fit value $w(0)=-1.114<-1$. Obviously, it can be shown that the fact
that $w(z)$ cross over the boundary of $w=-1$ in MCG model is
consistent with the results given by Refs. \cite{[14]} \cite{[35]}
\cite{[36]}, where $w(z)$ crossing -1 is first found using SNe data.
Furthermore, we obtain the $1\sigma$ confidence level of $w(0)$,
$-0.946\leq w(0)\leq-1.282$. The possibility of $w(0)>-1$
cann$^{,}$t be excluded in $1\sigma$ level. At last, it can be seen
that the cosmological constant model (i.e.,$w(z)=-1$) is not in
$1\sigma$ confidence contour of the best fit dynamical $w(z)$.

\section{The preferred cosmological model}

~~~~It is interesting to ask which model of an accelerating
universe is preferred by recently observed data over many models.
We also want to know how well the MCG model fits the recently
observed datasets as compared to other models. We make use of the
values of $\chi^{2}_{min}$ and the objective Alaike Information
Criterion (AIC)
 to solve the questions above.

On the basis of the description in section 3,
 we obtain the values of $\chi_{min}^{2}$ by minimizing the $\chi^{2}_{total-prior}$ in the
corresponding models, where $\Omega_{0m}$ is treated as a free
parameter with a Gaussian prior in the range $\Omega_{0m}=0.29\pm
0.07$ \cite{[2]} for the $\Lambda$CDM model and model-independent
cases,
 and the results are listed in Table 1. It
can be seen that the
 MCG model has the smallest $\chi^{2}_{min}$ value.  Table 1 contains the
best fit parameters corresponding to the different models.

 \begin{table}[ht]
\begin{center}
\begin{tabular}{|p{4.9cm}|p{1.3cm}|p{8.4cm}|} \hline
 case model                        & $\chi^{2}_{min}$ & Best fit parameters       \\ \hline
 $\Lambda$CDM                    & 162.302             & $\Omega_{0m}=0.286$ \\ \hline
 $w$=constant                   & 159.962  &            $\Omega_{0m}=0.288$, $w=-0.870$   \\ \hline
 $w(z)=w_{0}+w_{1}z$           & 159.765  &          $\Omega_{0m}=0.292$, $w_{0}=-0.893$, $w_{1}=0.009$
\\ \hline
 $w(z)=w_{0}+\frac{w_{1}z}{1+z}$        & 158.635   &  $\Omega_{0m}=0.289$,  $w_{0}=-1.041$, $w_{1}=0.751$
                                    \\\hline
$w(z)=w_{0}+\frac{w_{1}z}{(1+z)^{2}}$ & 157.715     &   $\Omega_{0m}=0.282$,  $w_{0}=-1.314$, $w_{1}=3.059$ \\
\hline
  $ w(z)=\frac{1+z}{3}\frac{A_{1}+2A_{2}(1+z)}{X}-1$
                                     & 159.067          & $\Omega_{0m}=0.292$, $A_{1}=-0.302$, $A_{2}=0.188$ \\ \hline
 GCG
                                  & 159.444      & $\Omega_{0b}=0.041$, $A_{s}=0.678$, $\alpha=-0.136$   \\ \hline
 MCG                                & 157.276         &
 \begin{small}
 $\Omega_{0b}=0.041$, $B=-0.085$, $B_{s}=0.822$,
                                                     $\alpha =1.724$
 \end{small}                                                     \\ \hline
\end{tabular}
\end{center}
\caption{\small{The values of $\chi^{2}_{min}$, and best fit model
parameters against the model }}
\end{table}

 In cosmology the AIC was first used by Liddle \cite{[37]},  and then in subsequent papers \cite{[38]} \cite{[39]}.
 It is defined as
\begin{equation}
AIC=-2\ln {\cal L}(\hat{\theta} \mid data)_{\max}+2K,\label{17}
\end{equation}
where ${\cal L}_{max}$ is the highest likelihood in the model with
the best fit parameters $\hat{\theta}$, $K$ is the number of
estimable parameters ($\theta$) in the model. The term $-2\ln {\cal
L}(\hat{\theta}\mid data)$ in Eq.(\ref{17}) is called $\chi^{2}$ and
it measures the quality of model fit, while the term $2K$ in
Eq.(\ref{17}) interprets model complexity.  For more details about
AIC please see Refs. \cite{[23]} \cite{[38]} \cite{[39]} \cite{[40]}
\cite{[41]} .

In what follows, we will estimate which model is the better one for
all the models in Table 2. The value of AIC has no meaning by itself
for a single model and only the relative value between different
models are physically interesting. Therefore, by comparing several
models the one which minimizes the AIC is usually considered the
best, and denoted by AIC$_{min}$$=$min$\{$ AIC$_{i}$,
$i=1,...,N$$\}$, where $i=1,...,N$ is a set of alternative candidate
models. The relative strength of evidence for each model can be
obtained by calculating the likelihood of the model ${\cal
L}(M_{i}\mid data)\propto $ exp $(-\bigtriangleup_{i}/2)$, where
$\bigtriangleup_{i}=$ AIC$_{i}-$AIC$_{min}$ over the whole range of
alternative models. The Akaike weight $w_{i}$ is calculated by
normalizing the relative likelihood to unity and corresponds to
posterior probability of a model. The evidence for the models can
also be judged by the relative evidence ratio
$\frac{w_{i}}{w_{j}}=\frac{{\cal L} (M_{i}\mid data)}{{\cal
L}(M_{j}\mid \text{data})}$. If model $i$ is the best one, the
relative evidence ratio gives the odds against the model. The rules
for judging the AIC model selections are as follows: when $0\leq
\bigtriangleup_{i}\leq 2$
 model $i$ has almost the same support from the data
as the best model, for $2\leq \bigtriangleup_{i}\leq 4$, model $i$
is supported considerably less and with $\bigtriangleup_{i} >10$
 model $i$ is practically irrelevant.

 \begin{table}[ht]
\begin{center}
\begin{tabular}{|p{5.2cm}|p{1.9cm}|p{1.9cm}|p{1.9cm}|p{1.9cm}|} \hline
 case model            &  AIC     &$\bigtriangleup _{i}$ & $w_{i}$ &Odds  \\ \hline
$\Lambda$CDM                  &  164.302   & 0.587  &0.149      & 1.342 \\
\hline $w$=constant           & 163.962    & 0.247  &0.177     & 1.130
\\ \hline
$w(z)=w_{0}+w_{1}z$           &  165.765   & 2.050  &0.072      & 2.778\\
\hline
 $w(z)=w_{0}+\frac{w_{1}z}{1+z}$ &164.635  &0.920  & 0.126     & 1.587\\\hline
$w(z)=w_{0}+\frac{w_{1}z}{(1+z)^{2}}$
                               &  163.715  & 0      & 0.200     & 1 \\\hline
 $ w(z)=\frac{1+z}{3}\frac{A_{1}+2A_{2}(1+z)}{X}-1$
                               & 165.067   &1.352   & 0.102     &1.961  \\ \hline
 GCG
                              & 165.444    &1.729   &0.084      & 2.381 \\ \hline
 MCG                          &  165.276   & 1.561  & 0.091     & 2.198 \\ \hline
\end{tabular}
\end{center}
\caption{\small{The value of AIC, Akaike difference, Akaike
weights $w_{i}$ and odds against the model}}
\end{table}

Thus based on the values of $\chi^{2}_{min}$ of all models  in
Table 1, the evidence of the AIC can be calculated. We find that
the best model is the one following
$w(z)=w_{0}+\frac{w_{1}z}{(1+z)^{2}}$ in terms of its AIC value.
Taking it as a reference, we calculate the differences between the
models by using the AIC differences $\bigtriangleup _{i}$, Akaike
weights $w_{i}$ and odds against alternative models. Table 2 gives
the calculating results. Note that the model selection provides
quantitative information to judge the "strength of evidence", not
just a way to select only one model. From Table 2 it is easy to
see that, the MCG model has almost the same support from the data
as the best model, because the value of $\bigtriangleup_{i}$ for
it is in the range 0-2. Furthermore, it can be shown that the
recent observational data supports all of the models in Table 2
except for the case of $w(z)=w_{0}+w_{1}z$ since the value of
$\bigtriangleup _{i}$ for this case is
 a little bigger than 2. It
  has a less support from recent observations.  On the other hand,
  we can see that the MCG model is favored by observational data more than GCG model
 according to the Akaike
weights $w_{i}$ in Table 2. Finally, the odds indicates the
difference between MCG model and the best one is 2.198 to 1.

\section{Conclusion}
~~~~In summary, the constraints on the MCG model, proposed as a
candidate of the unified dark matter-dark energy scenario, has been
studied in this Letter. We obtained the best fit value of the three
 parameters ($B$,$B_{s}$,$\alpha$) in the MCG model
 (-0.085,0.822,1.724).
 Meanwhile, it is easy to see that
the best fit $w(z)$ can cross -1 as it evolves with the redshift
$z$, and the present best fit value $w(0)=-1.114<-1$. Furthermore,
it is shown that the $1\sigma$ confidence level of $w(0)$ is
$-0.946\leq w(0)\leq-1.282$, and the possibility of $w(0)>-1$
cann$^{,}$t be excluded in $1\sigma$ level. We can see that the
cosmological constant model (i.e.,$w(z)=-1$) is not in $1\sigma$
confidence contour of the best fit dynamical $w(z)$.  Finally, in
order to find the status of MCG scenario in a large number of
cosmological models, we compared the MCG model with other seven
popular ones offering explanation of current acceleration of the
universe in terms of the values of $\chi^{2}_{min}$ and AIC
quantity. We find that, as the quantity $\chi^{2}_{min}$ measures
the quality of model fit, the MCG model is preferred  by recent
observational data because of its a small minimum $\chi^{2}$
value. On the other hand, it is shown that the MCG model has a
slightly high value of AIC due to its many parameters. However,
according to the rules of judgment of the AIC model selection, we
conclude that recently observed data supports the MCG model as
well as other popular models, because the value of
$\bigtriangleup_{i}$ for it is in the range 0-2 relative to the
best model. In addition, the result of study shows that the recent
observational data equivalently supports all of the models in
Table 2 except for the case of $w(z)=w_{0}+w_{1}z$. We expect the
new probers such as SNAP and Planck surveyor can provide more
accurate data and further explore the nature of dark energy.

\textbf{\ Acknowledgments } The research work is supported by NSF
(10573003), NSF (10747113), NSF (10573004), NSF (10703001), NSF
(10647110), NBRP (2003CB716300) and DUT(893326) of P.R. China.


\begin{thebibliography}{*}
\bibitem{[1]}   A.G. Riess, et. al, Astron. J. 116 (1998) 1009, astro-ph/9805201.

\bibitem{[2]}   D.N. Spergel, et. al, Astrophys. J. Suppl. 148
(2003)  175, astro-ph/0302209.

\bibitem{[3]}    A.C. Pope, et. al, Astrophys. J. 607 (2004) 655, arXiv:astro-ph/0401249.

\bibitem{[4]}    S. Weinberg, Mod. Phys. Rev. 61 (1989)  527.

\bibitem{[5]}  B. Ratra, P.J.E. Peebels, Phys. Rev. D. 37 (1988) 3406.

\bibitem{[6]}   R.R. Caldwell, M. Kamionkowski,  N.N. Weinberg,  Phys. Rev. Lett. 91 (2003) 071301, astro-ph/0302506.

\bibitem{[7]}    A.Y. Kamenshchik, U. Moschella, V. Pasquier,   Phys. Lett. B 511 (2001)  265, gr-qc/0103004.

\bibitem{[8]}    B. Feng, X.L. Wang, X.M. Zhang,  Phys. Lett. B 607 (2005) 35, astro-ph/0404224.

\bibitem{[9]}   M. Li, Phys. Lett. B 603 (2004) 1, hep-th/0403127.

\bibitem{[10]}   A.G. Riess, et. al,  Astrophys. J. 607 (2004) 665, astro-ph/0402512.

\bibitem{[11]}   A.R. Cooray, D. Huterer, Astrophys. J. 513 (1999) L95, astro-ph/9901097.

\bibitem{[12]}    E.V. Linder, Phys. Rev. Lett. 90 (2003) 091301, astro-ph/0208512.

\bibitem{[13]}    H.K. Jassal,  J.S. Bagla, T. Padmanabhan,   Mon. Not. R. Astron. Soc. 356
(2005) L11, astro-ph/0404378.

\bibitem{[14]}   U. Alam, et. al,   Mon. Not. R. Astron. Soc. 354 (2004) 275, astro-ph/0311364.

\bibitem{[15]}   H.B. Benaoum, hep-th/0205140.

\bibitem{[16]}    U. Debnath, A. Banerjee, S. Chakraborty, Class. Quantum Grav. 21 (2004)  5609,  gr-qc/0411015.

\bibitem{[17]}   L.P. Chimento, R. Lazkoz, Phys. Lett. B 615 (2005) 146, astro-ph/0411068.

\bibitem{[18]}   Y. Wu, et al., Mod. Phys. Lett. A 22 (2007) 11.

\bibitem{[19]}   X.Z. Li, D. J. Liu, Chin. Phys. Lett. 22 (2005).

\bibitem{[20]}   A.G. Riess et al.,  astro-ph/0611572.

\bibitem{[21]}  Y. Wang, P. Mukherjee, Astrophys. J. 650 (2006) 1, astro-ph/0604051.


\bibitem{[22]}   D.J. Eisenstein et al., Astrophys. J. 633
(2005) 560, astro-ph/0501171.

\bibitem{[23]}   M. Biesiada, J. Cosmol. Astron. Phys. 0702 (2007) 003, astro-ph/0701721.

\bibitem{[24]}   J.R. Bond, G. Efstathiou, M. Tegmark,  Mon. Not. R. Astron. Soc. 291 (1997) L33, astro-ph/9702100.

\bibitem{[25]}  D.N. Spergel et al., Astrophys. J. Suppl. 170 (2007) 377, astro-ph/0603449.

\bibitem{[26]}   $\O$. Elgar$\o$y, T. Multam$\stackrel{..}{a}$ki, astro-ph/0702343

\bibitem{[27]}  D.J. Eisenstein, W. Hu,   Astrophys. J. 496
(1998) 605, astro-ph/9709112.

\bibitem{[28]}   U. Alam, V. Sahni, Phys. Rev. D 73(2006) 084024, astro-ph/0511473.

\bibitem{[29]} W.L. Freedman et al., Astrophys. J. 553 (2001) 47, astro-ph/0012376.

\bibitem{[30]}  D. Kirkman et al., Astrophys. J. Suppl. 149 (2003) 1, astro-ph/0302006.

\bibitem{[31]}  C.B. Netterfield et al., Astrophys. J. 571 ( 2002) 604, astro-ph/0104460.

\bibitem{[32]} C. Pryke et al., Astrophys. J. 568 (2002) 46, astro-ph/0104490.

\bibitem{[33]}  S.W. Allen et al.,  Mon. Not. R. Astron. Soc. 353  (2004) 457, astro-ph/0405340.

\bibitem{[34]}  M. Goliath et al., Astron. Astrophys. 380 (2001) 6, astro-ph/0104009.

\bibitem{[35]} D. Huterer,  A. Cooray, Phys. Rev. D 71 (2005) 023506, astro-ph/0404062.

\bibitem{[36]}  T.R. Choudhury, T. Padmanabhan, Astron. Astrophys. 429 (2005) 807, astro-ph/0311622.


\bibitem{[37]}   A.R. Liddle,  Mon. Not. R. Astron. Soc. 351
(2004) L49, astro-ph/0401198.

\bibitem{[38]}  W. Godlowski, M. Szydlowski, Phys. Lett. B 623
(2005) 10, astro-ph/0507322.

\bibitem{[39]}  M. Szydlowski, W. Godlowski, Phys. Lett. B 633
(2006) 427, astro-ph/0509415.

\bibitem{[40]} L.X. Xu, C.W. Zhang, H.Y. Liu, Chin. Phys.
Lett. 24 (2007) 2459.

\bibitem{[41]}  M. Szydlowski, A. Kurek, A. Krawiec, Phys. Lett. B
642 (2006) 171, astro-ph/0604327.


\end{thebibliography}
\end{document}